\documentclass{rmaa}

\usepackage{paralist}
\usepackage{graphicx}
\usepackage{psfrag,color}
\title{Long-term photometric and spectroscopic observations of the near-contact binary KR Cygni\thanks{Based on 
  observations collected at T\"{U}B\.ITAK National Observatory (Antalya, Turkey).}}

\author{
E. Sipahi,\altaffilmark{1} 
C. \.{I}bano\v{g}lu\altaffilmark{1}
and \"{O}. \c{C}ak{\i}rl{\i},\altaffilmark{1}}

\altaffiltext{1}{Ege University, Science Faculty, Department of Astronomy and Space Sciences, 35100 Bornova, \.{I}zmir, TURKEY}

\shortauthor{Sipahi et al.}
\shorttitle{KR Cygni}
\fulladdresses{
\item Ege University, Science Faculty, Astronomy and Space Sciences Dept., 35100 
Bornova, \.{I}zmir, Turkey. $e-mail$: esin.sipahi@mail.ege.edu.tr}

\listofauthors{E. Sipahi,  \"{O}. \c{C}ak{\i}rl{\i} and C. \.{I}bano\v{g}lu}
\indexauthor{Sipahi, E.}
\indexauthor{\c{C}ak{\i}rl{\i}, \"{O}}
\indexauthor{\.{I}bano\v{g}lu, C.}

\abstract{We present the multi-color, five-year light curves and the first radial velocities of the near-contact binary system KR Cyg. We derived 
the masses of the components as 2.88$\pm$0.20 M$_{\odot}$ and 1.26$\pm$0.07 M$_{\odot}$ and the radii as 2.59$\pm$0.06 R$_{\odot}$ and 1.80$\pm$0.04 R$_{\odot}$. Analyses of the UBVR light curves and the radial velocities indicate that none of the components exactly fill their corresponding Roche lobes. We have calculated the distance to the system of KR\,Cyg as {411$\pm$12} pc using the observed apparent UBV magnitudes and the bolometric corrections for the component stars. We also searched for the empirical determination of albedo and effective temperature of the cooler, less massive star of KR\,Cyg, and of two similar 
near contact binaries AK\,CMi, and DO\,Cas. The residuals between the observed and computed fluxes are {\bf attributed to the effect of mutual illumination which heats the surface layers of the illuminated star and does vary not only its bolometric albedo but also its limb-darkening coefficient and gravity-brightening exponent.} The analysis of the light curves shows that the effective albedos are generally smaller than that expected from an envelope of convective star, being mostly departed from the theoretical value at the B passband. As the reflected light diminishes the effective temperature and, therefore, the luminosity of the irradiated star increase.  The observed bluer U-B colors during primary minimum are attributed to the effects of mutual irradiation and multiple scattering processes which may alter several characteristics of these systems.
}
\resumen{}
\addkeyword{ stars: binaries: close}
\addkeyword{ binaries: eclipsing}
\addkeyword{ binaries: general}
\addkeyword{ binaries: spectroscopic}
\addkeyword{ stars: individual: KR\,Cyg, AK\,CMi, DO\,Cas}
\begin{document}
\maketitle

\section{Introduction}                                                                                                               \label{sec:intro}
The light variability of KR\,Cyg (HD\,333645, $V=9^{m}.19$) was discovered by \citet{Sch31a, Sch31b, Sch31c}. The light curves obtained by photographic 
observations were published by \citet{Nek45} and \citet{Wac48} and the visual observations by \citet{Las36}, \citet{Gap53} and \citet{Tse54}. Both visual
 and photographic observations showed that the light variations were originated from the eclipses. The light curves were very similar to those of Algol-type 
binaries. The first photoelectric observations were made by  \citet{Vet65}. He obtained blue and yellow light curves of the system. The passbands used in the
 observations were very similar to those Johnson's B and V filters. A preliminary solution of the light curves leads him to the conclusion that the system 
consists of a B9 and an F5 main-sequence stars. He classified the system as a $\beta$ Lyrae type by visual inspection of the light curves. He also called 
attention some peculiarities in the course of the light curve. Later on Vetesnik's two-color light curves were analysed by \citet{Wil80} using a contemporary 
method, i.e.  the Wilson-Devinney code \citep{Wil71}, henceforth WD. The main difference between the two solutions is the primary star's radius which is fifteen 
per cent larger in the later than that found in the former analysis. In addition, \citet{Wil80} use an effective temperature for the primary star corresponding to 
a B7 star. The observations in the primary minimum of Vetesnik's light curves were re-analysed by \citet{Aln85} using the Fourier techniques in the frequency
 domain. They find that, in contrary to the older estimates, more massive star has smaller radius than its less massive companion.

\citet{Sip04} published three-color photometric observations and the resultant light curves of KR\,Cyg. The light curves were analysed individually
and simultaneously by the WD code. They determined a photometric mass-ratio to be 0.43 and suggested that the cooler, less massive component is nearly
filling its corresponding Roche lobe. Recently, one of us \citep{Sip12} collected all available times of light minimum and studied orbital period of 
the system. She arrived at a result that the differences between the times of observed and calculated (O-C) show a periodic oscillation with an amplitude 
of 0.001 days and a period of about 76 years. This periodic change was attributed to a hypothetical third component.

\citet{Sha90} defined a group of close binary stars named near-contact binaries (NCB) in which both components fill or nearly fill their critical Roche 
lobes. KR Cyg is included in the list of NCBs. According to his definition the stars of the near-contact binaries are near enough to each other to have 
strong proximity effects like W UMa type systems but are not in contact. As it is known the mass and energy exchange are being occurred between the 
components in the contact systems. This energy exchange is resulted in both components to come to nearly the same effective temperature. The main difference between the near-contact and contact binaries is the non-existence of the energy exchange through
the neck of material connecting the components in the near-contact systems. The non-existence of energy exchange in the case of NCB allows a large temperature difference between the components. Some NCBs show photometric asymmetries such that the system is brighter at the first quarter than that at the second one. This asymmetry is attributed to the primary star being one hemisphere, facing at phase about 0.25, is hotter than that at phase 0.75, or some material is obscuring the surface of the primary visible at phase 0.75. Therefore the NCBs are assumed to be lying in key evolutionary states of
close binary systems. \citet{Sha94} renewed the list of NCBs in which over 130 systems are given. He proposes that the NCBs are precursors of 
the A-type W UMa contact binaries.
 
The main aim of this study is to analyse the 1999 and 2000 BVR, and 2003, 2004, and 2005 UBVR light curves and the first radial velocities obtained 
by us. We derive the absolute physical parameters of the components and discuss light curve peculiarities by taking into account the mutual irradiation 
and therefore the changing in the value of albedo of the irradiated star for four passbands.

\section{Observations }
\subsection{Photometric observations}
The observations were made with the High-Speed Three-Channel Photometer attached to the 48 cm Cassegrain telescope at 
Ege University Observatory. HD 191398  and HD 333664, respectively, are taken as the comparison and check stars. These 
stars are nearly the same apparent magnitude and spectral type with the variable star. Moreover, they are very closely 
located with the variable on the plane of sky which make the effect of the atmospheric extinction on the differential 
magnitudes almost negligible. In Table 1 the coordinates, apparent visual magnitudes and spectral types of the variable
 and the comparison stars are given. The observations were made using the BVR passbands in 1999 and 2000, and UBVR 
passbands in 2003, 2004 and 2005. Though the comparison star is so close to the variable, the differential 
magnitudes, in the sense variable minus comparison, are corrected for atmospheric extinction. The nightly extinction
coefficients were derived from the observations of the comparison stars in each passband.

\begin{table*}
\centering
\footnotesize
\begin{minipage} {12 cm}
\caption{The coordinates, apparent visual magnitudes and the spectral types of the stars observed.}
\begin{tabular}{|l|c|c| c| c|}
\hline
\hline
\textbf{Star}	& \textbf {Alpha}	&\textbf {Delta}    &\textbf{V (mag)}	&	\textbf {Sp}	\\
\hline					
\textbf{KR Cyg}	    & 20 09 06	&30 33 01     & 9.23	&	A0V	\\
\textbf{HD 191398} 	& 20 08 39  &30 20 15     & 9.01	&   A0V \\
\textbf{HD 333664}	& 20 09 18  &30 13 39     & 9.58    &   A0	\\
\hline
\end{tabular}
\end{minipage}
\end{table*}
\smallskip

\subsection{Spectroscopic observations}
Optical spectroscopic observations of KR\,Cyg were obtained with the Turkish Faint Object Spectrograph Camera (TFOSC)
attached to the 1.5 m telescope on 4\,nights (July-August, 2011) under good seeing conditions. Further details on the telescope and the 
spectrograph can be found at http://www.tug.tubitak.gov.tr. The wavelength coverage of each spectrum was 4000-9000 \AA~in 12 orders, with a 
resolving power of $\lambda$/$\Delta \lambda$ 7\,000 at 6563 \AA~and an average signal-to-noise ratio (S/N) was $\sim$120. We also obtained a 
high S/N spectrum of the $\alpha$\,Lyr (A0 V) and HD\,27962 for use as templates in derivation of the radial velocities \citep{Nidever}. 

The electronic bias was removed from each image and we used the 'crreject' option for cosmic ray removal. Thus, the resulting 
spectra were largely cleaned from the cosmic rays. The echelle spectra were extracted and wavelength calibrated by
using Fe-Ar lamp source with help of the IRAF {\sc echelle} \citep{Tonry_Davis} package. 
 
The stability of the instrument was checked by cross correlating the spectra of the standard star against each other using 
the {\sc fxcor} task in IRAF. The standard deviation of the differences between the velocities measured using {\sc fxcor} and the 
velocities in \citet{Nidever} was about 1.1 km\,s$^{-1}$.

\section{Analyses}
\subsection{Radial velocities}
To derive the radial velocities for the components of binary system, the four TFOSC spectra of the eclipsing binary, obtained near the quadratures,  were 
cross-correlated against the spectrum of $\alpha$\,Lyr on an order-by-order basis using the {\sc fxcor} package in IRAF. The majority 
of the spectra showed two distinct cross-correlation peaks in the quadrature, one for each component of the binary. Thus, both peaks were 
fitted independently in the quadrature with a $Gaussian$ profile to measure the velocity and errors of the individual components. If the two peaks appear 
blended, a double Gaussian was applied to the combined profile using {\it de-blend} function in the task. For each of the four observations we, 
then, determined a weighted-average radial velocity for each star from all orders without significant contamination by telluric absorption
features. Here we used as weights the inverse of the variance of the radial velocity measurements in each order, as reported by {\sc fxcor}.

We adopted a $two-Gaussian$ fit algorithm to resolve cross-correlation peaks near the first and second quadratures when spectral lines are 
visible separately. Fig.\,1 shows examples of cros-correlations obtained by using the largest FWHM at nearly first and second quadratures. The 
two peaks, correspond to each component of KR\,Cyg. The stronger peaks in each CCF correspond to the more luminous 
component which has a larger weight into the observed spectrum.

\begin{figure*}
\includegraphics[width=12cm]{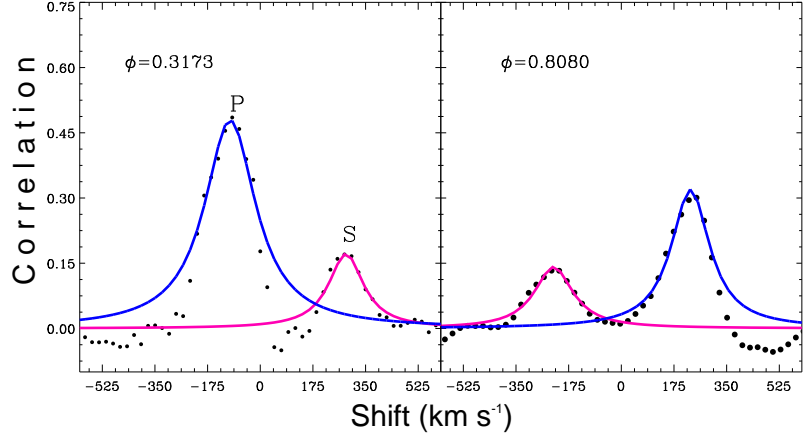}
\caption{Sample of CCFs between KR Cyg and the RV template spectrum around the first and second quadrature.}
\end{figure*}

\begin{table}
\centering
\begin{minipage}{85mm}
\caption{Heliocentric radial velocities of KR\,Cygni. The columns give the heliocentric Julian date, the
orbital phase, the radial velocities of the two components with the corresponding standard deviations.}
\begin{tabular}{@{}ccccccccc@{}c}
\hline
HJD 2400000+ & Phase & \multicolumn{2}{c}{Star 1 }& \multicolumn{2}{c}{Star 2 } 	\\
             &       & $V_p$                      & $\sigma$                    & $V_s$   	& $\sigma$	\\
\hline
55751.36069 &	0.8080	&  73.1    & 	11.8	& -290.8 &	11.3\\
55796.38592 &	0.0826	& -92.3	   &  7.8	& 77.7	  &11.2\\
55835.36321 &	0.2013	& -150.4 &	6.8	& 188.3 &	9.5\\
55835.46128 &	0.3173	& -144.6 &	9.9	& 184.1 &	9.9\\
\hline \\
\end{tabular}
\end{minipage}
\end{table}

The heliocentric RVs for the primary (V$_p$) and the secondary (V$_s$) components are listed in Table\,2, along with the dates 
of observation and the corresponding orbital phases computed with the new ephemeris given by \citet{Sip12}. The velocities in this 
table have been corrected to the heliocentric reference system by adopting a radial velocity of 9.5 km\,s$^{-1}$ for the template 
star $\alpha$\,Lyr. The RVs listed in Table\,2 are the weighted averages of the values obtained from the cross-correlation 
of orders \#4\#, \#5\#, \#6\# and \#7\# of the target spectra with the corresponding order of the standard star spectrum. The weight 
$W_i = 1/\sigma_i^2$ has been given to each measurement. The standard errors of the weighted means have been calculated on 
the basis of the errors ($\sigma_i$) in the RV values for each order according to the usual formula (e.g.\citet{toping}). The $\sigma_i$ 
values are computed by {\sc fxcor} according to the fitted peak height, as described by \citet{Tonry_Davis}.

The measured radial velocities of both stars are plotted in Fig. 2 against the orbital phase.  First we analysed the radial velocities for the 
initial orbital parameters. We used the orbital period held fixed and computed the eccentricity of the orbit, systemic velocity and 
semi-amplitudes of the RVs. The results of the analysis are as follows: $e$=0.001$\pm$0.001, i.e. formally consistent with a circular 
orbit as expected, $\gamma$= -43$\pm$3 km\,s$^{-1}$, $K_1$=110$\pm$4 km\,s$^{-1}$ and $K_2$=251$\pm$7 km\,s$^{-1}$. Using 
these values we estimate the projected orbital semi-major axis and mass ratio as: $a$sin$i$=6.028$\pm$0.127 R$_{\odot}$ and 
$q=\frac{M_2}{M_1}$=0.438$\pm$0.017.

\begin{figure}
\includegraphics[width=10cm=0]{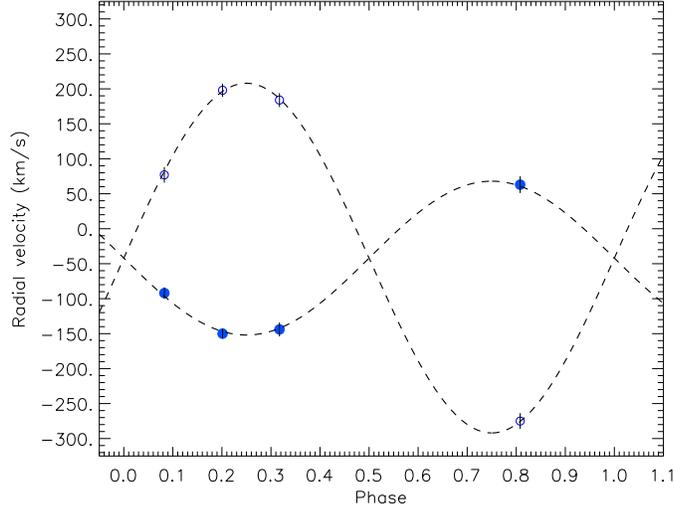}
\caption{\label{fig:CCF} Radial velocities of the components.}
\end{figure}

\subsection{Effective temperature of the primary star}
The spectral type of a star can be derived from the multi-color photometry and spectroscopy, or both. Unfortunately, UBV measurements of the system 
are not available. Therefore we observed the system with the UBV standard stars some nights and obtained the standard apparent visual magnitude and 
colors at mid-secondary eclipse as: V=9.231$\pm$5, U-B=0.101$\pm$15, and B-V=0.156$\pm$7 mag. The $Q$-parameter of the Johnson's wide-band 
photometry, defined as $Q$=(U-B) -0.72 (B-V) \citep{Joh53}, is independent of interstellar reddening for early type stars. Using the observed (U-B) and (B-V) 
colors at secondary eclipse we calculated the $Q$-parameter as -0.011 for the hotter star. Then, we estimated the intrinsic color of (B-V)$_0$ = -0.021 mag
from the tables given by  \citet{Hov04}. Since the light contribution of the secondary component does not exceed a few hundreths of magnitudes in the B 
and V passbands its contribution to the B-V color at maximum light may be ignored. This intrinsic color  corresponds to an A0$\pm$1 main-sequence star. The average effective 
temperature deduced from the calibrations of \citet{bohm},  \citet{dejager},  \citet{flower} and \citet{drill} is about 9\,810 K. The spectral-type uncertainty 
leads to an uncertainty of about 150 \,K in the effective temperature of the primary star. The difference between the observed and intrinsic B-V color, i.e. the 
interstellar reddening, is about 0.177 mag. 

KR Cyg was observed by \citet{hild75} in the intermediate-band photometric system. One of their observations falls into the primary eclipse. Excluding this 
observation we calculated the colors of $u-b$=1.555$\pm$0.011, $b-y$=0.111$\pm$0.010 and $c$=0.970$\pm$0.022 mag. Taking E($b-y$)=0.73E(B-V) we computed 
the intrinsic color of ($b-y$)$_0$=-0.018$\pm$0.010 mag which corresponds to an A0 star. The infra-red magnitudes J=9.278$\pm$0.022, H=9.169$\pm$0.022 and 
K=9.098$\pm$0.018 are given in the $2MASS$ catalog \citep{cutri}. The infrared colors are calculated as J-H=0.109$\pm$0.031 and H-K=0.071$\pm$0.028 which
correspond to a reddened B9$\pm$2 star. The spectral type of the primary star obtained both intermediate and infrared photometric systems is consistent 
what we derived from the UBV photometry.      

\subsection{Analyses of the light curves}
As we stated in \S 2.1 we have multi-color light curves of KR\,Cyg obtained with the same instrumentation.
We used the recent version of the eclipsing binary light curve modelling algorithm of \citet{Wil71} (with updates), as 
implemented in the {\sc phoebe} code of Pr{\v s}a \& Zwitter (2005). The code needs some input parameters, such as the effective 
temperature of the hotter star and the mass-ratio of the system. These two parameters are the key-parameters for the solution 
which should be estimated before beginning the light curve modelling. The effective temperature of the hotter 
star, eclipsed in the primary minimum, has been estimated using the photometric indices as  9\,810$\pm$150 K. 
The mass ratio, $q$=$M_2$/$M_1$, is very important parameter in the light curve analysis, because the WD code is based on 
Roche geometry which is sensitive to this quantity. The mass ratio of 0.438$\pm$0.017 determined from the radial velocities 
was kept as a fix value.  A preliminary estimate for the effective temperature of the cooler component is made using the depths of the eclipses.

The logarithmic limb-darkening coefficients, $x_1$  and $x_2$, and bolometric limb-darkening coefficients, $x_{bol}$ and $y_{bol}$, were 
determined from tables by Van Hamme (1993) for the primary and secondary components, respectively, taking into account the effective temperatures and the 
wavelengths of the observations. Standard values of bolometric albedos \citep{rucinski}, and the gravity-darkening coefficients \citep{lucy} for radiative and 
convective  envelopes were used. The rotational velocities of the components are assumed to be synchronous with the orbital one. The adjustable parameters 
in the light curves fitting were the orbital inclination $i$, the surface potentials $\Omega_1$ and $\Omega_2$, the effective temperature of secondary 
T$_2$, the luminosity of the primary star $L_1$.  
 
Using a trial-and-error method, we obtained a preliminary set of parameters which represent the observed UBVR light curves. The code was set in Mode--2, a 
detached configuration, with coupling between luminosity and effective temperature and no constraints on the potentials. We used a simple reflection 
treatment as $MREF=1$ and  $NREF=1$ at the beginning of iterations.
 
The iterations were carried out automatically until convergence and a solution was defined as the set of parameters for which the differential corrections 
were smaller than the probable errors. After a few iterations convergence was obtained. The final results of the five-year multi-color light curves' 
simultaneous analysis of KR Cyg are listed in Table 3. $r_1$ and $r_2$ in this table are the mean fractional radii of the hotter and cooler components, 
respectively. The uncertainties assigned to the adjusted parameters are the internal errors provided directly by the Wilson-Devinney code. We also ran 
models for KR\,Cyg in mode Mode--5, the secondary component filling its Roche lobe, in  Mode--4, the primary filling its Roche lobe and in Mode--3, both 
components fill their Roche lobes. Convergence is obtained in  Modes--4 and 3 but with larger sum of squared residuals.  
 
The computed light curves (continuous line) are compared with the observations in four passbands in Figure 3. While there is a satisfactory fit between 
the observed and computed lights  for the U passband the binary model does not represent the observed light curves successfully in the B, V, and R 
passbands. The residuals, in the sense observed minus calculated, are plotted in Fig.4 for a better visibility.  The differences between the model 
and the observed light curves are clearly seen at both maxima in three passbands and particularly in the secondary eclipse in  B and R passbands. As pointed out 
by \citet{Sha90} all of the NCBs do not show O'Connell effect, one maximum is higher than the other. If a NCB shows an asymmetry in the light curve, 
the first maximum is brighter than the second maximum. This difference is generally attributed to the secondary component, in which the visible surface 
at phase 0.25 is brighter than the other half. While the asymmetries reach to 5\% of the total light, the secondary components in these systems 
contribute less than 5\% of the system's total light. Therefore, it is suggested that the visible hemispheres of the hotter components at the 
orbital phase of 0.75 are either cooler or obscured by a cool material.

\begin{figure*}
\includegraphics[width=10cm=0]{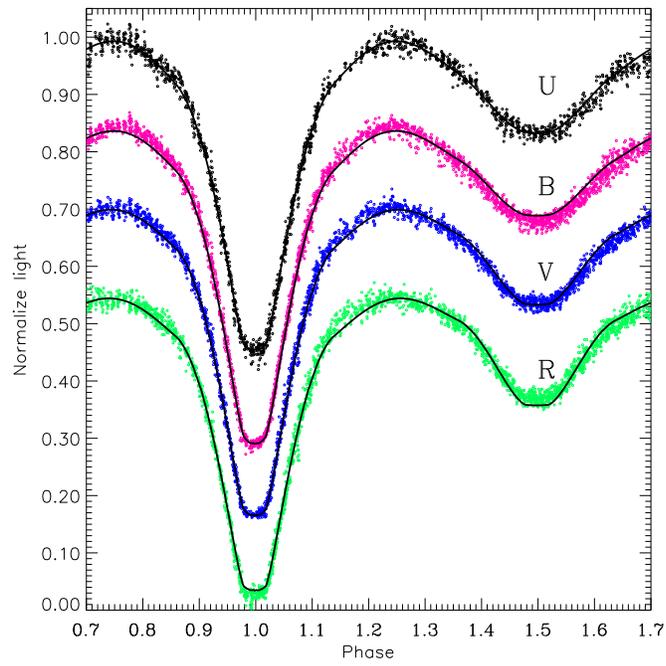}
\caption{The observed five-year UBVR flux scale light curves and computed light curves for KR\,Cyg.}
\end{figure*}

\begin{table}
  \caption{Fitted parameters for the five-year multi-color light curves of KR\,Cyg.}
  \label{parameters}
  \begin{tabular}{lccccccccccc}
  \hline
   Parameter & U   &   B  &    V  &     R                 \\
   \hline
$\lambda$~(\AA)& 3500&4300&	5500&6700  	\\ 
N$_{obs}$	&1752 &2825 &2804 & 2825				\\  
$x_1$&0.620&0.714&0.611&0.773  	\\   
$x_2$&0.871&0.844&0.773&0.682  	\\   
$g_1$&1.0&1.0&1.0&1.0  	\\   
$g_2$&0.32&0.32&0.32&0.32  	\\   
$A_1$&1.0&1.0&1.0&1.0  	\\   
$A_2$&0.5&0.5&0.5&0.5  	\\   
$T_1$~(K)&9810&9810&9810&9810  	\\   
$q$=$M_2$/$M_1$&0.438&0.438&0.438&0.438  	\\ 
$L_{1}/(L_{1+2})$&0.9791$\pm$0.0005 &0.9727$\pm$0.0004&0.9467$\pm$0.0005&0.9179$\pm$0.0006  	\\
$i (^{\circ})$ 			& \multicolumn{4}{c} {87.00$\pm$0.02}  	 \\   
$T_{2}$ (K)				&  \multicolumn{4}{c} {5\,580$\pm$10} 	 \\
$\Omega_{1}$ 			& \multicolumn{4}{c} {2.8645$\pm$0.0012} \\
$\Omega_{2}$ 			& \multicolumn{4}{c} {2.8111$\pm$0.0008} \\     
$r_1$					& \multicolumn{4}{c} {0.4297$\pm$0.0002} \\
$r_2$					& \multicolumn{4}{c} {0.2989$\pm$0.0002} \\   
$\chi^2$				& \multicolumn{4}{c} {1.821}  			 \\
\hline \end{tabular}
\end{table}

In the case of KR\,Cyg the light curve is almost symmetric in all passbands. At phases of about 0.15 and 0.85 the observed fluxes are larger than 
those computed in the BVR passbands. In contrary, the observed fluxes are smaller than those computed at the secondary eclipse in the B, V and R passbands. The components 
in the NCBs are nearly touch each other but their photosphere are not in physical contact. A gaseous stream and an energy exchange  between 
the stars may be excluded. 

 \subsection{ Search for the empirical derivation of the albedo}
The value of the local emergent bolometric flux and energy balance in a distorted star are significantly dependent upon the assumed values of the
gravity darkening exponent and the bolometric albedo \citep{Raf80}. The reflection effect is used to describe the mutual irradiation of the facing 
hemispheres of the components in a close binary system. For the stars of intermediate temperatures of about 10 000K the incoming radiation is 
almost completely absorbed by atoms and ions in the irradiated atmosphere which alters the temperature structure and causes an increased 
emergent flux. For the stars of convective envelopes half of the incoming radiation is converted to bulk motions of the atmosphere gases 
\citep{Ruc69}. The theoretical value of bolometric albedo is assumed to be about 0.5 for the stars having convective atmospheres. Furthermore,  \citet{Vaz85} 
calculated bolometric reflection albedo for a particular reflecting  main-sequence star with a temperature of 4 500 K and presented the results 
as a function of angle of incidence and relative incident flux. As pointed out by \citet{Cra93} the effect of mutual irradiation will be strongest on the 
line-of-centers points of the stars in a close binary system. The stars themselves iterate this effect between one another, e.g., hotter star heats up 
the cooler that in turn affects the illumination back on the hotter star, and so on. The facing hemispheres of both components are heated up by multiple 
scattering process.       
 
\begin{figure} 
\begin{center}
\includegraphics[width=12cm=0]{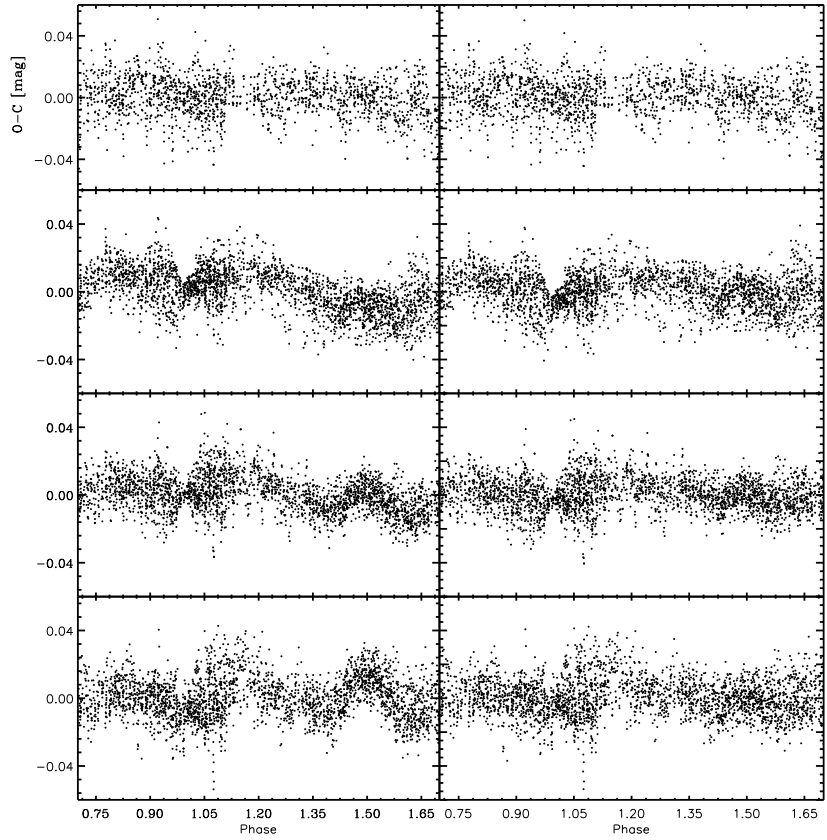}
\caption{\label{fig:CCF} Residuals, in light units,  for the two solutions of KR\,Cyg's UBVR light curves from top to bottom: $A_2$=0.5 (fixed, left 
panels) and $A_2$ is an adjustable parameter (right panels).}
\end{center} \end{figure}

\begin{table}
 \setlength{\tabcolsep}{2.5pt} 
  \caption{Absolute parameters of KR\,Cyg.}
  \label{parameters}
  \begin{tabular}{lcc}
  \hline
   Parameter 						& Primary							&	Secondary									\\
   \hline
   $a$ (R$_{\odot}$)				&\multicolumn{2}{c}{6.04$\pm$0.13}														\\
   $V_{\gamma}$ (km s$^{-1}$)		&\multicolumn{2}{c}{-43$\pm$5} 														\\
   
   $q$							&\multicolumn{2}{c}{0.438$\pm$0.017}													\\
   Mass (M$_{\odot}$) 				& 2.876$\pm$0.198 						&1.261$\pm$0.074									\\
   Radius (R$_{\odot}$) 			& 2.594$\pm$0.055 						&1.804$\pm$0.038									\\
   $\log~g$ ($cgs$) 				& 4.069$\pm$0.012 						&4.026$\pm$0.012									\\
   $T_{eff}$ (K)					& 9\,810$\pm$150						&5\,580$\pm$150      								\\
   $(vsin~i)_{obs}$ (km s$^{-1}$)	& 141$\pm$7							&108$\pm$9       									\\
   $(vsin~i)_{calc.}$ (km s$^{-1}$)	& 155$\pm$3							&108$\pm$2		    								\\
   $\log~(L/L_{\odot})$				& 1.750$\pm$0.032						&0.454$\pm$0.050       								\\
   $d$ (pc)						& \multicolumn{2}{c}{411$\pm$12}														\\
    $U$, $B$, $V$ (mag)		& \multicolumn{2}{c}{9.488$\pm$0.014, 9.387$\pm$0.005, 9.231$\pm$0.005}					\\
   $J$, $H$, $K_s$ (mag)		& \multicolumn{2}{c}{7.159$\pm$0.018, 7.221$\pm$0.027, 7.249$\pm$0.027}					\\
\hline  
\end{tabular}
\end{table}

\begin{table}
  \caption{The calculated effective albedos and effective temperatures for the cooler components of KR\,Cyg, AK\,CMi and DO\,Cas.}
  \label{parameters}
  \begin{tabular}{lccccccccccc}
  \hline
   Star &Parameter & U   &   B  &    V  &     R                 \\
   \hline
KR\,Cyg & $A_2$&0.435$\pm$0.023 &0.225$\pm$0.016 &0.241$\pm$0.012 &0.304$\pm$0.012   	\\   
KR\,Cyg & $T_2$&5600$\pm$27 &5985$\pm$14 &5653$\pm$12 &5336$\pm$12 	\\   
KR\,Cyg & $L_2$&0.022&0.041&0.057&0.070  	\\   
AK\,CMi & $A_2$&0.457$\pm$0.041 &0.132$\pm$0.038 &0.219$\pm$0.028 &---   	\\   
AK\,CMi & $T_2$&5658$\pm$60 &5880$\pm$45 &5677$\pm$32 &--- 	\\   
DO\,Cas & $A_2$&0.275$\pm$0.158 &0.377$\pm$0.127 &0.333$\pm$0.089 &---   	\\   
DO\,Cas & $T_2$&5441$\pm$147 &5615$\pm$105 &5427$\pm$74 &--- 	\\  
  \hline
  \end{tabular}
\end{table}

Therefore we attempted to estimate the values of  $g_2$ and  $A_2$ from analyses of well-defined multi-color light curves of the system. WD code allows 
us to make parallel solutions by solving any number of subsets of adjustable parameters in the same computer run. By adopting the theoretical values of 
the less sensitive parameters of limb-darkening coefficients, gravity-darkening exponents and the limb-darkening coefficients we derived the basic set of 
parameters $i$, $\Omega_{1}$, $\Omega_{2}$, $L_{1}$, $L_{2}$ and $T_{2}$ from the  analyses of multi-color light curves. The components of the 
system KR\,Cyg are nearly contact and they rotate faster than  100 km s$^{-1}$, and have a large temperature difference of about 4\,200 K. Therefore, the 
gravity-darkening exponent and the bolometric albedo for the components are expected to be different from the non-rotating and gravitationally undistorted 
stars. In the second step, we take gravity-darkening exponent, effective temperature, luminosity and effective albedo for the cooler component as adjustable 
parameters in the analysis of U, B, V, and R light curves separately. A few iterations indicate that there is no significant deviation in the gravity-darkening 
exponent between the theoretical and calculated values. The iterations are carried on excluding the gravity-darkening. After a few iterations a convergence 
is attained. The computed values of   $A_{2}$,and $T_{2}$ are given with their standard deviations in Table\,5. The computed $L_{2}$ values are also presented.   Following the formulation given by \citet{Cra93} and \citet{Cla01} we connect the bolometric albedo with the intrinsic and incident flux by

\begin{equation}
\sigma T_h^4=\sigma T_{eff}^4 + F_{incident}
\end{equation}

\begin{equation}
A_{F_r}=\sigma T_h^4 - T_{eff}^4
\end{equation}
where T$_{eff}$ is the effective temperature of the star without external flux, T$_{h}$ is the effective temperature after irradiation, F$_r$ is the external radial flux and A is the bolometric albedo. The external flux F$_r$ is depended on apparent radius, effective temperature of the irradiating star, and the cosine of the angle of incidence. The amount (1-A)F$_r$ is absorbed and the atmosphere of the irradiated star re-emit this excess energy later. Moreover, \citet{Kir76} and \citet{Kir82} proposed that not only the irradiated hemisphere of the star is heated in close binaries. The horizontal fluxes 
resulting from the incident radiation field can cause large circular currents over the irradiated stellar surfaces and may penetrate into the non-directly irradiated hemisphere.  As a result the irradiated star would increase its temperature and luminosity. The values given in Table 5 confirm this theoretical prediction. The empirically derived values of the $A_{2}$, and $T_{2}$ from
 the light curves are plotted versus the wavelengths of the observations in Fig.\,5. The derived values of $A_{2}$ are systematically 
 smaller than their theoretical values of about 0.5. Moreover, the deviations appear to be depended upon the wavelengths. The largest deviation occurs 
 about $\lambda$4350. While the albedo gets smaller the effective temperature of the illuminated star increases. The computed value of albedo in the U-passband is close to the theoretically expected value for convective envelopes. The light contribution of the secondary star to the total light is too low for this passband. {\bf The influence of the mutual illumination in close binary systems is not only seen on the bolometric albedos but also in other properties of the stars such as limb-darkening coefficients and gravity-darkening exponents. The reflection effect has a strong influence on limb-darkening coefficients, as already noticed by \citet{Vaz85}, \citet{Nor90}, \citet{Cla90}. Later on, the effect of illumination on the limb-darkening coefficients is studied numerically by \citet{Ale99} using the Uppsala Model Atmosphere code in convective line-blanketed atmospheres. Their results show that the limb-darkening coefficients of illuminated atmospheres are significantly different from the non-illuminated ones. They have concluded that the limb-darkening coefficients vary depending on the characteristics of external illumination such as the incidence angle, the amount of infalling flux, the temperature of the illuminating star. In addition, \citet{Alen99} showed that external illumination changes the gravity-brightening exponent of an illuminated star. They have proposed that the contact and semi-detached eclipsing binary systems should be better represented by illuminated atmospheres, since their components are closer to each other. Their numerical calculations show that the external illumination increases the values of gravity-brightening exponent roughly in proportional to the amount of the incident flux. The classical value of 0.32 for the convective atmospheres may be too small for binary systems with close components. }

We also solved the UBV light curves of AK\,CMi, a similar system to KR\,Cyg, obtained by  \citet{Sam98}. They estimated an effective temperature of 8\,510\,K 
for the hotter star. We derived a value of 10 150 K using the UBV measurements of the system at four orbital phases, at maxima and  in mid-eclipses. Applying 
the same procedure as in KR\,Cyg we obtained  $A_{2}$, and $T_{2}$ for the cooler component. They are listed in Table 5 and plotted in the middle panel of 
Fig.\,5. Though the uncertainties are slightly larger both astrophysical parameters indicate very similar variations as in the cooler component of 
KR\,Cyg. DO\,Cas is also a NCB and its UBV light curves were published by \citet{Oh92}. We derived an effective temperature for the hotter star as 
8\,700\,K using the intermediate-band measurements by \citet{hild75} and the BVJHK magnitudes given in the SIMBAD database. Initial parameters were 
adopted from \citet{Siw10} who obtained spectroscopic mass-ratio and modelled the BV light curves. Their analysis resulted in near-contact 
configuration.  Applying the same procedure we derive  $A_{2}$ and $T_{2}$ for the irradiated star of DO\,Cas. The results are given in Table\,5 
and plotted in the bottom panels of Fig.5. The uncertainties of the parameters are very large, originated from the large scatters in the 
observational data, the effective albedos are smaller than the expected value of 0.5. 

\begin{figure}
\begin{center}
\includegraphics[width=9cm=0]{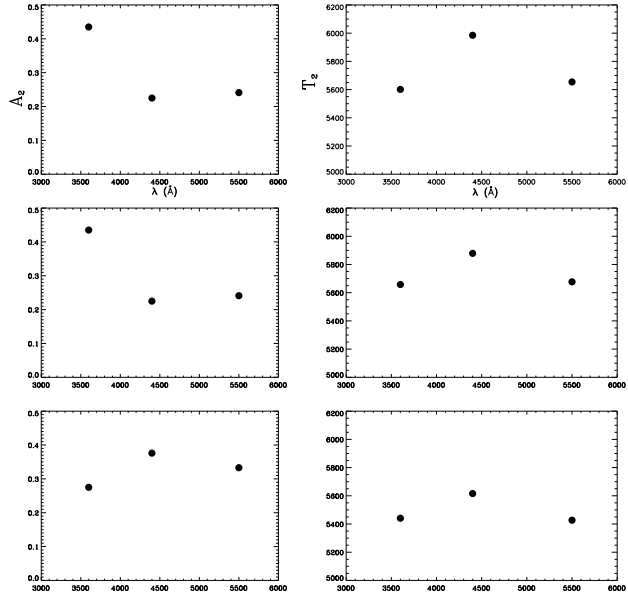}
\caption{\label{fig:CCF} The empirical values of the effective albedos (left panels) and temperatures (right panels) are plotted against the wavelengths for the cooler components of KR Cyg (top panels), AK CMi (middle panels) and DO Cas (bottom panels). }
\end{center}
 \end{figure}

In Fig.\,6 we plot the observed U-B and B-V colors for the systems  KR\,Cyg, AK\,CMi and DO\,Cas. The bluer color especially in the U-B just in and around primary eclipse is clearly seen. About 50 per cent of the apparent disk of the hotter component of KR\,Cyg is obscured by the cooler star in mid-primary eclipse. We, therefore, suggest that multiple scattering of the light between the components of close binaries with large temperature differences results
in bluer color, especially in the annular eclipse.

\begin{figure}
\begin{center}
\includegraphics[width=10cm=0]{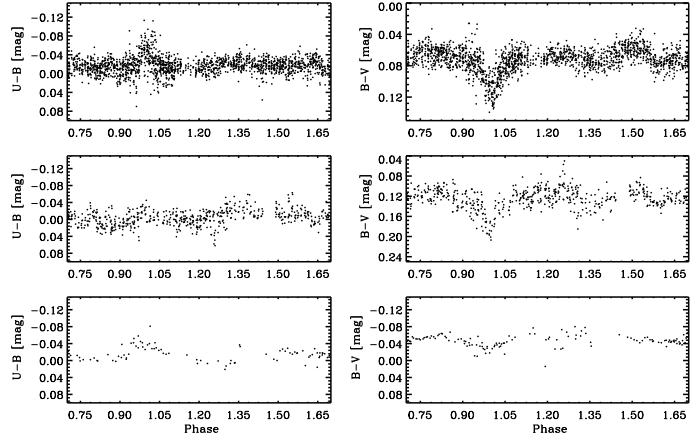}
\caption{\label{fig:CCF} The U-B and B-V colors against the orbital phase for KR Cyg (top panels), AK CMi (middle panels) and DO Cas (bottom panels). }
\end{center}
 \end{figure}

 \section{Conclusions }
We obtained the five-year UBVR light curves and the radial velocities of the near-contact binary KR\,Cyg. Analysing the multi-color 
data and measured radial velocities we obtained, for the first time, the absolute parameters of the components. Comparison of the components' 
radii with the corresponding Roche lobes indicate that the system appears to be near-contact but not in contact. Combining the results of 
light-and radial-velocity analyses we obtained the absolute physical parameters of the system. Both components locate on the main-sequence of 
the Hertzsprung-Russell diagram. Systematic behaviour of the residuals between the observed and fitted light curves is attributed to the effect 
of the mutual heating of the components, mainly in the case of the cooler star, of this near-contact system. Analyses indicate that the effective 
albedo and the effective temperature of the irradiated star are significantly altered. The empirically derived value of albedo is smaller than that 
expected for a convective star. While the albedo gets smaller the effective temperature increases. We also analyzed the UBV data, taken from 
literature, for the systems AK\,CMi and DO\,Cas. The changes of albedos versus the wavelength are very similar for AK\,CMi and KR\,Cyg. {\bf The analysis of eclipsing binary light curves involves a large number of parameters. The bolometric albedos, limb-darkening coefficients, gravity-brightening exponents are mostly kept fixed at the theoretical values. Nevertheless, external illumination is significant in close eclipsing binary systems which heats the surface layers of the illuminated star, thus alters its physical properties. }

\section*{Acknowledgment}
We thank to Turkish Scientific and Technical Research Council (T\"UB\.ITAK) for a partial support with Project number 109T708 
and EB{\.I}LTEM Ege University Science Foundation Project No:2011/B\.{I}L/007.We also thank to T\"{U}B{\.I}TAK National Observatory 
(TUG) for a partial support in using RTT150 and T100 telescopes with project numbers 10ARTT150-483-0, 11ARTT150-123-0 and 
10CT100-101. We also thank to responsible for the Bak{\i}rl{\i}tepe observing station for their warm hospitality. This research 
has been made use of the ADS and CDS databases, operated at the CDS, Strasbourg, France and T\"{U}B\.{I}TAK 
ULAKB{\.I}M S\"{u}reli Yay{\i}nlar Katalo\v{g}u.

{\bf....}

\end{document}